\documentclass{article}
\usepackage{setspace}
\usepackage{arxiv}

\usepackage[utf8]{inputenc} 
\usepackage[T1]{fontenc}    
\usepackage{hyperref}       
\usepackage{url}            
\usepackage{booktabs}       
\usepackage{amsfonts}       
\usepackage{nicefrac}       
\usepackage{microtype}      
\usepackage{graphicx}
\usepackage{natbib}
\usepackage{doi}
\usepackage{subfiles}
\usepackage{amsmath}
\usepackage{subcaption}
\usepackage{xcolor}
\usepackage{ulem}

\title{Image registration of 2D optical thin sections in a 3D porous medium: \\Application to a Berea sandstone digital rock image}

\author{%
  \href{https://orcid.org/0000-0003-2960-4601}{\includegraphics[scale=0.06]{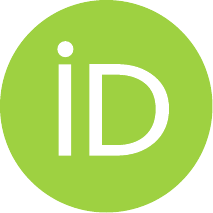}\hspace{1mm}Jaehong Chung\thanks{Corresponding author: jhchung1@stanford.edu}}\textsuperscript{1, *} \quad
  \href{https://orcid.org/0000-0001-5919-8734}{\includegraphics[scale=0.06]{orcid.pdf}\hspace{1mm}Wei Cai}\textsuperscript{2} \quad
  \href{https://orcid.org/0000-0003-1711-1850}{\includegraphics[scale=0.06]{orcid.pdf}\hspace{1mm}Tapan Mukerji}\textsuperscript{1,3} 
  \\[8pt] 
  \textsuperscript{1}Department of Geophysics, Stanford University, Stanford, CA, USA  \\[3pt]  
  \textsuperscript{2}Department of Mechanical Engineering, Stanford University, Stanford, CA, USA   \\[3pt]  
  \textsuperscript{3}Department of Energy Science and Engineering, Stanford University, Stanford, CA, USA  
}

\date{}

\renewcommand{\undertitle}{ }
\renewcommand{\shorttitle}{ } 

\hypersetup{
pdftitle={A template for the arxiv style},
pdfsubject={q-bio.NC, q-bio.QM},
pdfauthor={David S.~Hippocampus, Elias D.~Striatum},
pdfkeywords={First keyword, Second keyword, More},
}

\begin{document}
\maketitle

\begin{abstract}
This study proposes a systematic image registration approach to align 2D optical thin-section images within a 3D digital rock volume. Using template image matching with differential evolution optimization, we identify the most similar 2D plane in 3D. The method is validated on a synthetic porous medium, achieving exact registration, and applied to Berea sandstone, where it achieves a structural similarity index (SSIM) of 0.990. With the registered images, we explore upscaling properties based on paired multimodal images, focusing on pore characteristics and effective elastic moduli. The thin-section image reveals 50 \% more porosity and submicron pores than the registered CT plane. In addition, bulk and shear moduli from thin sections are 25 \% and 30 \% lower, respectively, than those derived from CT images. Beyond numerical comparisons, thin sections provide additional geological insights, including cementation, mineral phases, and weathering effects, which are not clear in CT images. This study demonstrates the potential of multimodal image registration to improve computed rock properties in digital rock physics by integrating complementary imaging modalities.
\end{abstract}


\section{Introduction}
\label{sec:Introduction}
Understanding pore-scale processes is essential for various subsurface applications, including groundwater flow \citep{zhu1999pore, bear2010modeling}, carbon dioxide sequestration \citep{chen2019inertial, chung2023generating, chung2024accelerating}, underground hydrogen storage \citep{hashemi2021pore, wang2023pore}, and radioactive waste disposal \citep{curti2019modelling, chung2019numerical}. Traditionally, these processes have been investigated through laboratory experiments using core plugs, which are time-consuming and challenging to reproduce. Recently, digital rock physics (DRP) utilizes high-resolution digital rock images to perform pore-scale analysis and simulate physical processes to compute effective properties \citep{andra2013digital_1, andra2013digital_2}. DRP provides a non-destructive and cost-effective alternative to conventional laboratory measurements \citep{berg2017industrial}. In particular, micro-CT imaging has become a widely used tool in DRP for computing effective properties, offering 3D rock microstructures while preserving sample integrity. It enables permeability estimation via pore-scale fluid flow simulations \citep{blunt2013pore} and elastic moduli estimation via micro-scale mechanical simulations \citep{zhu2017modeling, saxena2016estimating}. More recently, deep learning approaches have leveraged micro-CT images to accelerate the prediction of effective rock properties, improving computational efficiency \citep{liu2023hierarchical, ahmad2023homogenizing, chung2024prediction, chung2024predicting}.

Despite these advantages, three-dimensional CT images present several challenges. First, the resolution may not be sufficient to capture all pore structures \citep{blunt2017multiphase, saxena2019rock}. As resolution increases, the sample size that can be imaged decreases. For instance, a $2000^3$-voxel dataset at 1 $\mu m$ resolution represents a 2 mm$^3$ volume. Given the fixed total number of voxel, increasing the resolution by reducing the pixel size to 0.5 $\mu m$ halves the sample size to 1 mm$^3$. Since the studied volume must be at least as large as the representative elementary volume (REV) to derive meaningful upscaling properties, a coarser resolution is commonly chosen to image a larger sample exceeding the REV \citep{yoon2013nanopore, milani2016representative}. Consequently, typical micro-CT scans of rocks have resolutions ranging from 1 to 20 $\mu m$ for samples a few to a few tens of millimeters in size \citep{bazaikin2017effect, blunt2017multiphase, saenger2011digital}. However, submicron pores, grain roughness, small throat channels, and microcracks—features that significantly influence upscaled properties—often remain unobserved in micro-CT images but are detectable at higher resolutions \citep{bazaikin2017effect}. Second, micro-CT images do not provide explicit geological information. While grayscale intensity variations allow density-based differentiation between grains and pores, they do not distinguish mineral compositions \citep{niu2020digital, goldfarb2022predictive}. Additionally, cementation and weathering effects, which critically impact rock properties \citep{basu2009evaluation, cook2015effect}, are not explicitly visible in CT images . Third, the computed effective properties are subject to systematic errors due to the aforementioned resolution limitations and the inability to identify geological information. In particular, these issues can lead to porosity underestimation (by a factor of 0.5) and permeability overestimation (by a factor of 10) \citep{saxena2019rock}. Similarly, effective elastic moduli derived from digital images tend to be stiffer than those measured in laboratory tests \citep{saxena2019rock2}.

Additional imaging modalities can help overcome these limitations by providing complementary information. Thin-section microscope images offer submicron resolution (e.g., 0.65 $\mu m$/pixel \citep{araya2020deep}, 0.74 $\mu m$/pixel \citep{saxena2017estimating}) and reveal mineralogy, grain contacts, fine-scale pore structures, and weathering features \citep{saxena2021application}. However, thin-section images capture only 2D information, making it difficult to infer the broader 3D rock microstructure. Data-driven super-resolution techniques have attempted to fuse CT and thin-section data \citep{niu2020innovative, liu2022multiscale}, but they often assume that submicron features in thin-section images match those in the entire CT volume, which may not hold for heterogeneous or anisotropic rocks. If the thin section is not representative of the full 3D volume, reconstructed 3D features may be misinterpreted. Therefore, before applying unpaired image fusion techniques, it is essential to verify that the thin-section image is spatially aligned with the 3D CT volume.

\begin{figure}[htbp]
    \centering
    \includegraphics[width=\linewidth]{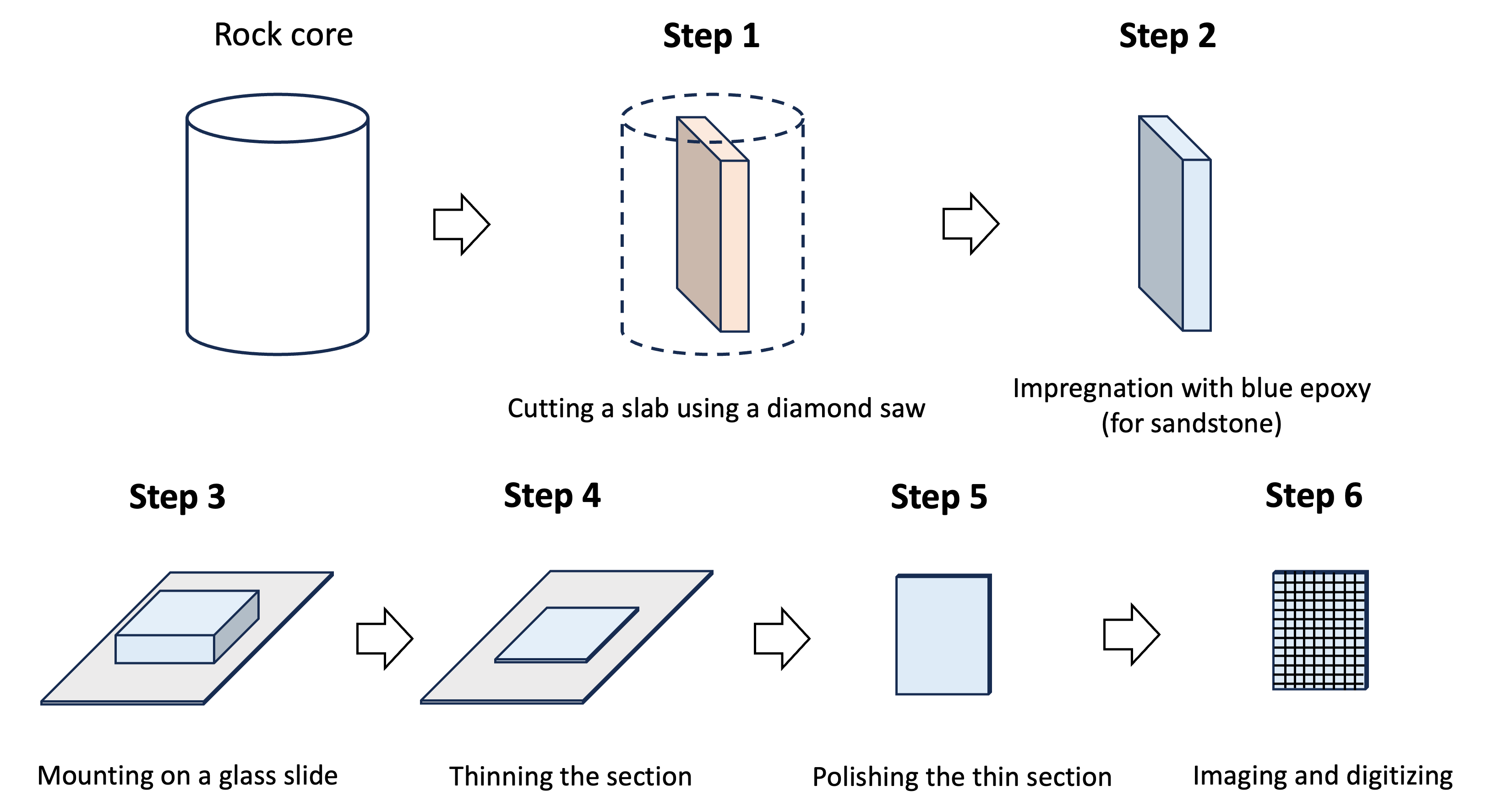}
    \caption{Thin section preparation process}
    \label{fig:TS_prep}
\end{figure}

A persistent challenge in multimodal imaging is the precise alignment of a 2D thin section within a 3D CT volume. Manual feature matching is time-consuming, prone to errors, and complicated by uncertainties introduced during thin section preparation. Figure \ref{fig:TS_prep} shows a general procedure for preparing thin-section digital rock images, which includes multiple steps such as sawing, polishing, and digitizing. Each of these steps introduces potential misalignment, complicating the precise spatial correlation between the thin section and the original rock sample. Consequently, accurate registration of thin-section and micro-CT images remains challenging, even with a known rough location, limiting the effective integration of dual-modality information. 

Image registration—which aligns images from different modalities into a single coordinate system—is widely used in fields such as medical imaging and automatic target recognition \citep{brown1992survey, zitova2003image}. In medicine, for example, physicians frequently combine CT scans (which depict bone structures) with MRI data (which highlight soft tissues) to track tumor growth or plan treatments, as each modality provides complementary information about the same anatomical region \citep{maurer1993review, maes2003medical}. Similarly, multimodal image registration in geoscience can improve the accuracy of effective rock properties that align with laboratory measurements by integrating CT and thin-section data. However, geological applications of image registration remain largely unexplored, particularly for aligning a 2D thin section within a 3D CT volume, which introduces 3D rotational challenges.

To address this challenge, we propose an optimization approach that employs cross-correlation analysis with 3D rotations to locate the best-matched interface between thin-section and micro-CT images. This approach aims to (i) reduce manual alignment efforts, (ii) explore the advantages of paired multimodal images in improving digital rock-based computed effective properties.

To demonstrate the versatility of the proposed method, we apply it to both synthetic 3D porous media, serving as controlled testbeds for validation, and Berea sandstone samples. These applications highlight the method’s effectiveness across a range of 3D porous structures. Furthermore, with the two registered images of Berea sandstone, we analyze differences in pore characteristics and upscaling properties by comparing images from the same location in two different modalities. Our results indicate that precise registration can help correct image-based simulated properties, improving the accuracy of image-based rock property estimates. Additionally, this study provides a foundation for future data-driven super-resolution techniques that rely on spatially consistent multimodal imaging.

\section{Method}
\label{sec:Method}
\begin{figure}[htbp]
    \centering
    \includegraphics[width=\linewidth]{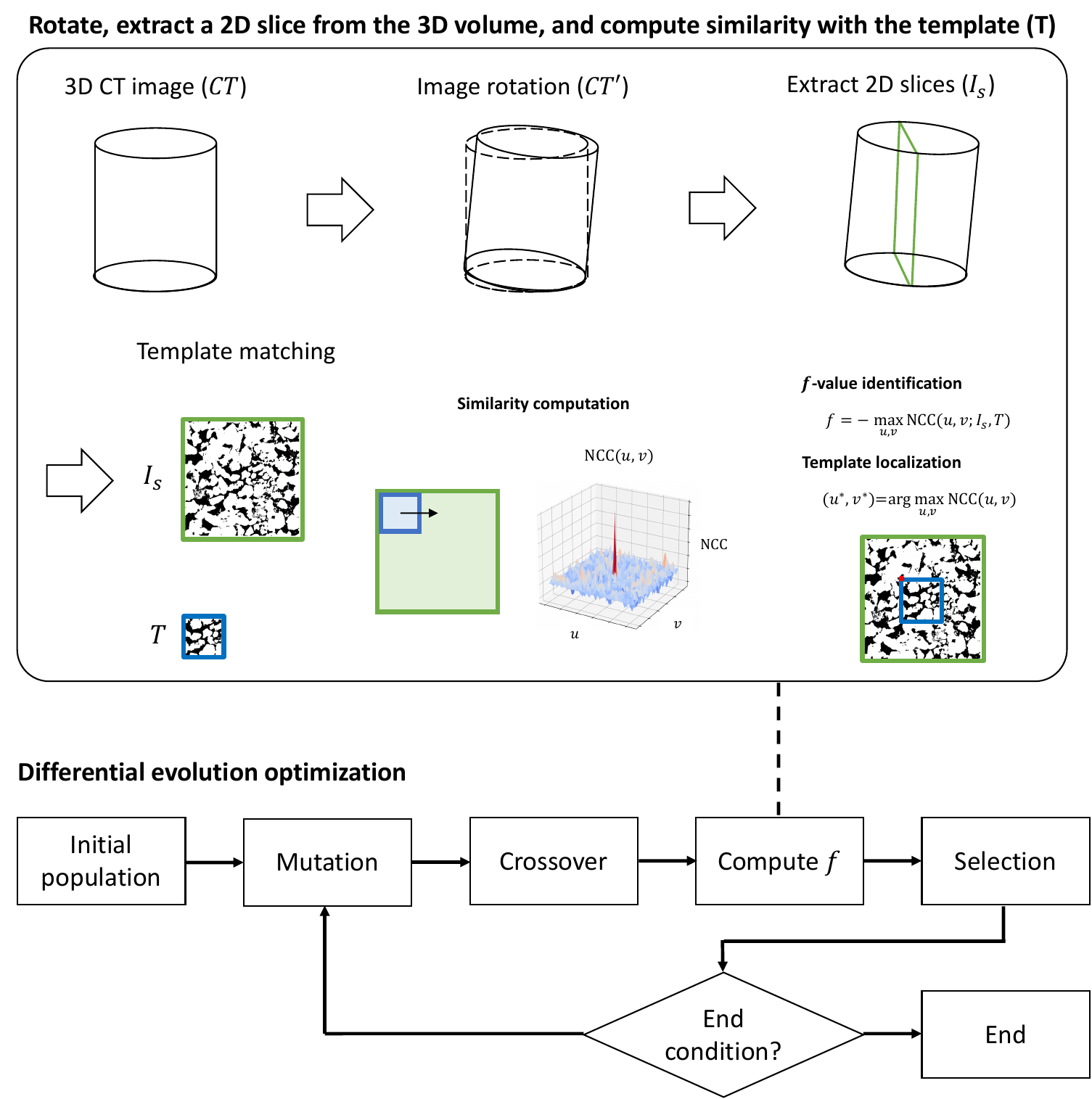}
    \caption{Workflow for 2D image registration within a 3D CT volume via optimization. The upper section represents the process of applying rotations to the 3D CT scan, extracting a 2D slice, and computing similarity with the template image. The lower section shows the differential evolution optimization process, which iteratively updates the transformation parameters to minimize the objective function \( f \), derived from the similarity evaluation in the upper section.}
    \label{fig:workflow}
\end{figure}

We developed a systematic approach to register a 2D thin section within a 3D image. Figure \ref{fig:workflow} shows the overall procedure of the image registration, which consists of four steps: 
(1) performing a linear transformation of the 3D CT scan, 
(2) extracting a 2D image from the rotated 3D volume, 
(3) applying template matching to evaluate the similarity between the extracted image and the thin section image, and 
(4) repeating (1)--(3) to find the optimal registration via an optimization algorithm, specifically the differential evolution (DE) algorithm used in this study.

Before performing this workflow, a preprocessing step may be necessary depending on the dataset. In general, when two images differ in their feature spaces, filters can be applied to make the features more coherent, and the pixel size can be matched by rescaling \citep{brown1992survey, zitova2003image}. In this section, we assume that the two images have the same resolution and feature space to explain the general procedure of our registration workflow, as preprocessing steps may vary depending on the dataset. In the section \ref{sec:sub_result2_BereaSandstone}, we will further discuss the preprocessing steps required for Berea sandstone digital images. Here, we present the general procedure for registering any 2D image to a 3D CT volume.

First, a linear transformation is applied to the binarized 3D CT volume. This transformation involves rotating around the principal axes of the 3D image. The rotation angles \(\alpha\), \(\beta\), and \(\gamma\) are applied sequentially around the \(x\)-, \(y\)-, and \(z\)-axes using the following rotation matrices:

\begin{equation}
    R_x(\alpha) = 
    \begin{bmatrix}
        1 & 0 & 0 \\
        0 & \cos(\alpha) & -\sin(\alpha) \\
        0 & \sin(\alpha) & \cos(\alpha)
    \end{bmatrix}, 
    \quad
    R_y(\beta) = 
    \begin{bmatrix}
        \cos(\beta) & 0 & \sin(\beta) \\
        0 & 1 & 0 \\
        -\sin(\beta) & 0 & \cos(\beta)
    \end{bmatrix},
    \quad
    R_z(\gamma) = 
    \begin{bmatrix}
        \cos(\gamma) & -\sin(\gamma) & 0 \\
        \sin(\gamma) & \cos(\gamma) & 0 \\
        0 & 0 & 1
    \end{bmatrix}.
\end{equation}

The composite rotation matrix is:
\begin{equation}
    R = R_z(\gamma)\, R_y(\beta)\, R_x(\alpha).
\end{equation}

We obtain the rotated CT volume by:
\begin{equation}
    CT' = CT \cdot R,
\end{equation}
where \(CT\) is the binarized 3D volume image, and \(CT'\) is the rotated image. After rotation, a 2D slice is extracted from the rotated image at slice index \(s\) along the original \(x\)-axis:
\begin{equation}
    I_s = CT'(\alpha, \beta, \gamma)\Big|_{x = s}.
\end{equation}

We then employ a template-matching algorithm to quantify the similarity between the extracted CT section \(I_s\) and the thin section image \(T\). Specifically, we compute the normalized cross-correlation (NCC) over every possible location of \(T\) within \(I_s\). To ensure zero-mean data, we first subtract the mean intensity from both images:

\begin{equation}
    T' = T - \bar{T}, 
    \quad 
    I_s' = I_s - \bar{I_s}.
\end{equation}

The normalized cross-correlation (NCC) map is defined as:

\begin{equation}
    \text{NCC}(u, v) = 
    \frac{
        \sum_{p, q} \bigl[I_s'(p + u,\, q + v)\, T'(p, q)\bigr]
    }{
        \sqrt{\Bigl[\sum_{p, q} (I_s'(p + u,\, q + v))^2\Bigr]\,
        \Bigl[\sum_{p, q} (T'(p, q))^2\Bigr]}
    },
\end{equation}

where \( (u, v) \) denote the offsets applied to the reference image \( I_s \), and \( (p, q) \) range over the valid region of the template \( T \), satisfying \( 0 \leq p < w \) and \( 0 \leq q < h \), where \( w \) and \( h \) denote the width and height of \( T \), respectively.

The peak value of the NCC map indicates the highest similarity between \( I_s \) and \( T \), corresponding to the optimal alignment. Thus, for a given rotation and slice index, we can determine the best placement of the thin section \( T \) within the trial section \( I_s \). We formulate the objective function for optimization as follows:

\begin{equation}
    f(\alpha, \beta, \gamma, s) = 
    - \max_{u, v} \text{NCC}(u, v \;\mid\; I_s, T),
\end{equation}
\begin{equation}
    (\alpha^*, \beta^*, \gamma^*, s^*) = 
    \arg \min_{\alpha, \beta, \gamma, s} f(\alpha, \beta, \gamma, s).
\end{equation}

Differential evolution (DE) is an evolutionary algorithm to minimize this objective function \citep{storn1997differential}. DE initializes a population of parameter vectors given the bounds of each of the four parameters. The initial population at generation~0 is

\begin{equation}
    \theta_i^{(0)} = \bigl[\alpha,\; \beta,\; \gamma,\; s\bigr]_i^{(0)},
    \quad i = 1, \ldots, N_P,
\end{equation}

where \(N_P\) is the population size. In each generation \(t\), every target vector \(\theta_i^{(t)}\) evolves into \(\theta_i^{(t+1)}\) by undergoing mutation, crossover, and selection. In the mutation step, the best current solution \(\theta_{\mathrm{best}}^{(t)}\) (the one with the lowest \(f\)) and two other distinct population members \(\theta_{r1}^{(t)}\) and \(\theta_{r2}^{(t)}\) are used to form a donor vector:

\begin{equation}
    d_i^{(t)} = 
    \theta_{\mathrm{best}}^{(t)} 
    + F\,\bigl[\theta_{r1}^{(t)} - \theta_{r2}^{(t)}\bigr],
\end{equation}

where \(F\) is the differential weight. In the crossover step, the donor \(d_i^{(t)}\) and the target \(\theta_i^{(t)}\) are combined to produce a trial vector \(t_i^{(t)}\):

\begin{equation}
    t_{i,j}^{(t)} \;=\;
    \begin{cases}
        d_{i,j}^{(t)}, & \text{if } r_j \le CR \text{ or } j = j_{\mathrm{rand}},\\
        \theta_{i,j}^{(t)}, & \text{otherwise},
    \end{cases}
\end{equation}

where \(CR\) is the crossover rate, \(r_j\) is a random number in \([0,1]\), and \(j_{\mathrm{rand}}\) ensures that at least one parameter is inherited from the donor. Finally, in the selection step, the trial vector \(t_i^{(t)}\) replaces \(\theta_i^{(t)}\) if it yields a lower objective function:

\begin{equation}
    \theta_i^{(t+1)} = 
    \begin{cases}
        t_i^{(t)}, & \text{if } f\bigl(t_i^{(t)}\bigr) \,\le\, f\bigl(\theta_i^{(t)}\bigr),\\
        \theta_i^{(t)}, & \text{otherwise}.
    \end{cases}
\end{equation}

DE iterates these steps until it reaches the maximum number of iterations or until improvement in the objective function is below a specified tolerance. In this study, the tolerance is set to \(\epsilon = 1\times 10^{-5}\), and the maximum number of iterations is 2000. Thus, our workflow combines template matching with normalized cross-correlation and differential evolution to systematically register a 2D image within a 3D volume.

\section{Results}
\label{sec:Results}
\subsection{Verification of the thin section registration based on known location}
The registration approach was validated using a synthetic porous medium with a predefined 2D plane location. The validation dataset consisted of a 300-voxel cubic sphere pack \citep{dataset_spherepack_2019}. A 2D plane was extracted after applying an arbitrary rotation of $\left[5,\, -3,\, 4\right]$ degrees along the x-, y-, and z-axes, with the section number set to 155. This predefined location served as the ground truth for validation.

\begin{figure}[htbp]
    \centering
    \includegraphics[width=\linewidth]{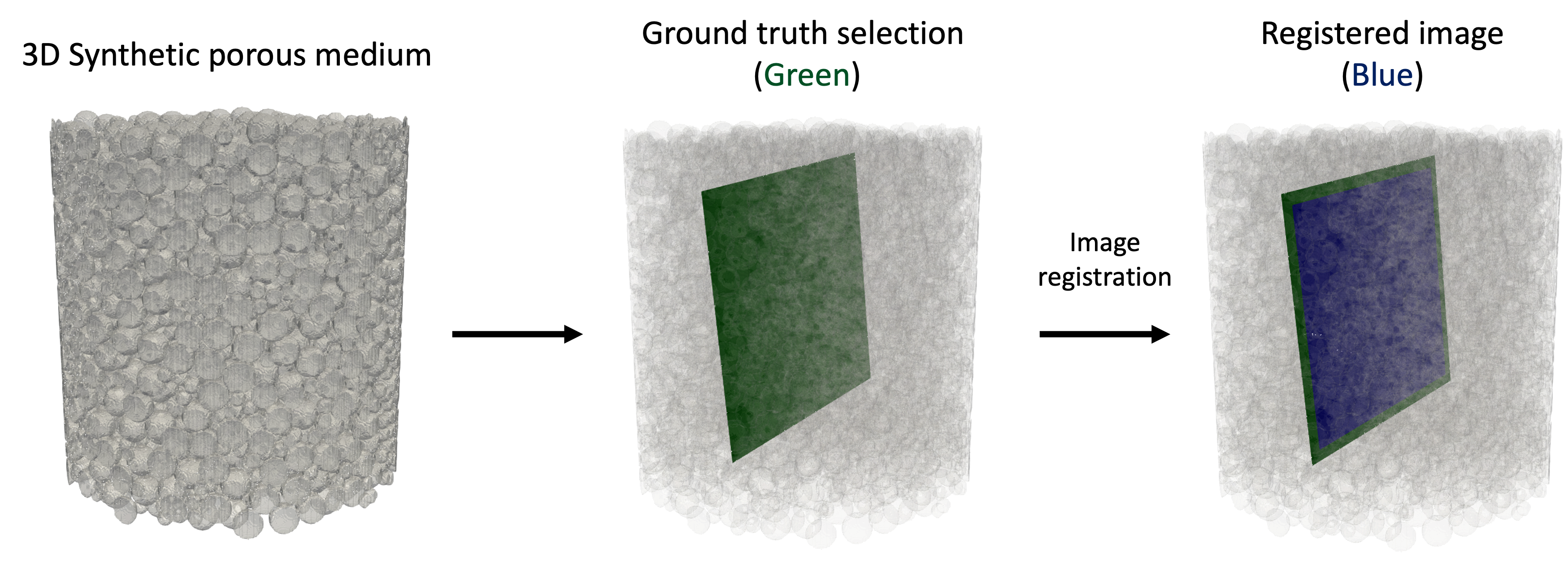}
    \caption{Validation of the 2D image registration approach in a 3D synthetic porous medium. (left) The initial 3D sphere pack porous medium. (middle) A known 2D plane selected as the ground truth (template) for the image registration (green). (right) The registered 2D image after image registration (blue), overlaid with the ground truth for validation.}
    \label{fig:validation_process}
\end{figure}

For image registration, the rotation angle bounds were set to 0 - 10$^\circ$, and the section number was constrained to the range 1/4 - 3/4 of the cubic size (i.e., 75 - 225). The registration algorithm was then applied. Figure \ref{fig:validation_process} shows the validation process, including the original porous medium, the extracted ground truth plane as a template, and the registered image after optimization.

\begin{figure}[htbp]
    \centering
    \includegraphics[width=0.8\linewidth]{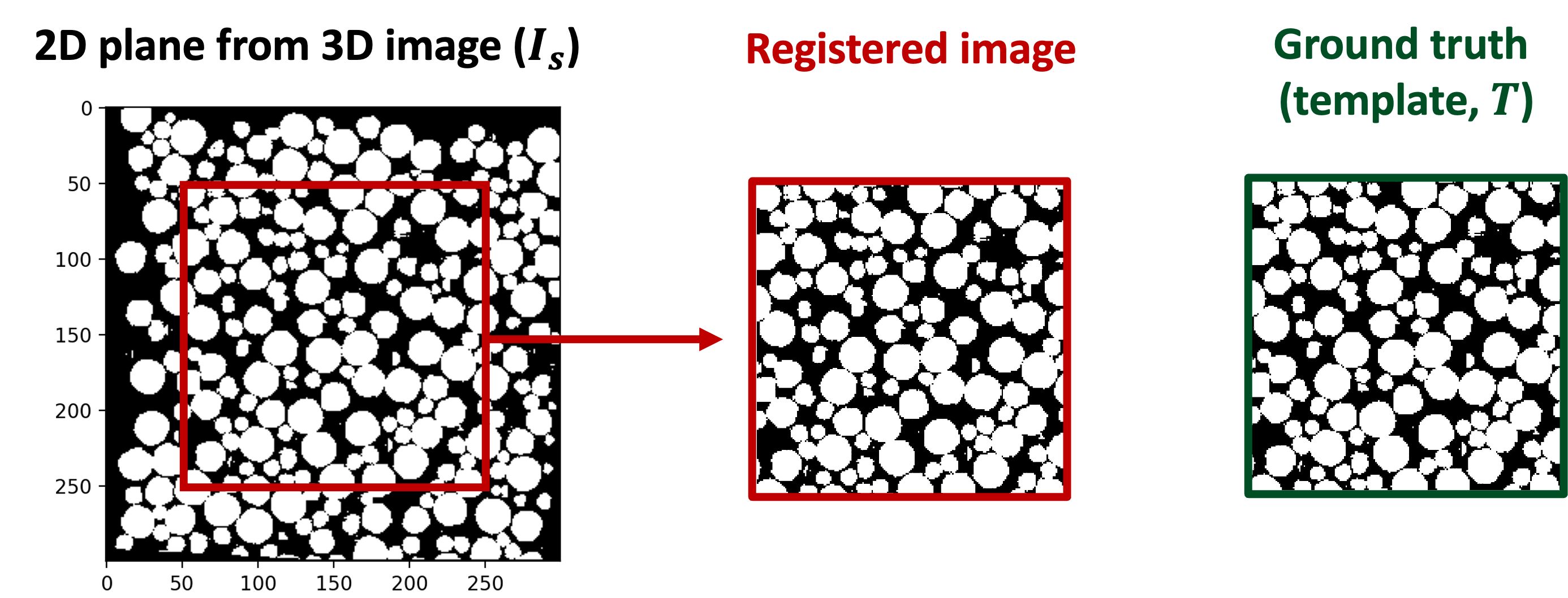}
    \caption{Validation case: comparison between the registered image extracted from the 3D volume ($I_s$) and the ground truth template ($T$).}
    \label{fig:validation_superimposed}
\end{figure}

\begin{figure}[htbp]
    \centering
    \includegraphics[width=0.4\linewidth]{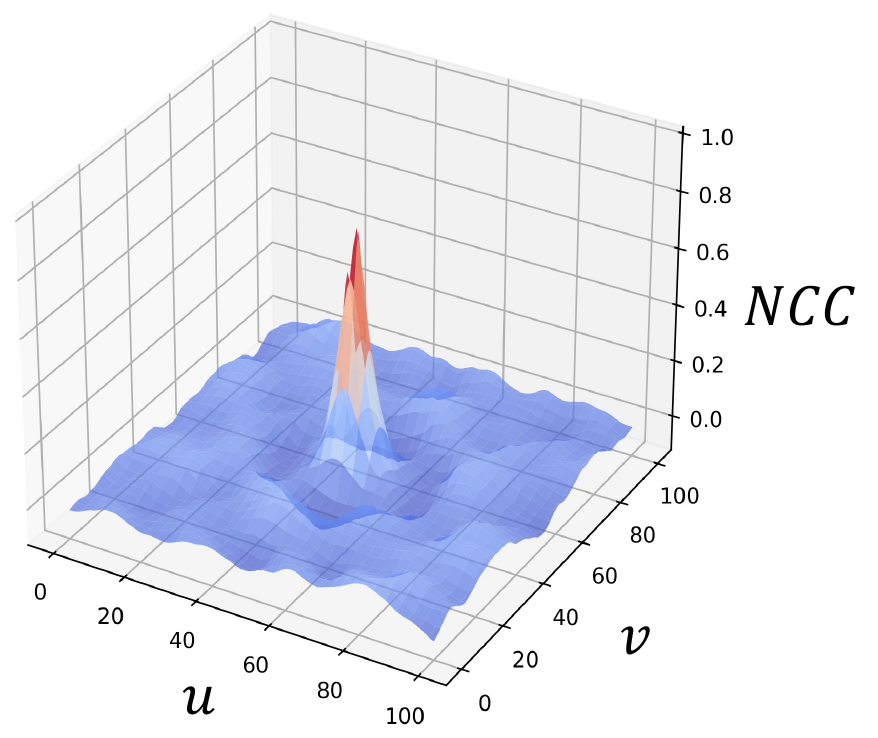}
    \caption{Validation case: normalized correlation coefficient maps between the 2D plane from 3D image ($I_s$) and ground truth template ($T$) in the Figure \ref{fig:validation_superimposed}}
    \label{fig:3D_correlation_map_validation}
\end{figure}

The registration results indicate that the error in x-, y-, and z-angles was less than 0.001 degrees, and the estimated section number exactly matched the ground truth. Table \ref{tab:validation_registration_results} presents the detailed results. Figure \ref{fig:validation_superimposed} shows the superimposed registered section with the ground truth, confirming the valid alignment. Figure \ref{fig:3D_correlation_map_validation} shows the peak correlation coefficient between the ground truth and registered image was 0.99999, indicating a high degree of coherence. These results demonstrate that our approach can accurately register a 2D image within a 3D volume.

\begin{table}[htbp]
    \centering
    \caption{Validation case: comparison between the ground truth and registered image after image registration.}
    \label{tab:validation_registration_results}
    \begin{tabular}{l|ccc|c}
        \hline
        & \multicolumn{3}{c|}{rotation angles [degrees]} & section number \\
        & x & y & z &  \\
        \hline
        ground truth      & 5.000  & -3.000  & 4.000  & 155 \\
        registered image  & 4.999  & -3.000  & 3.999  & 155 \\
        \hline
    \end{tabular}
\end{table}

\subsection{Thin section registration in Berea sandstone CT image}
\label{sec:sub_result2_BereaSandstone}
\begin{figure}[htbp]
    \centering
    \includegraphics[width=\linewidth]{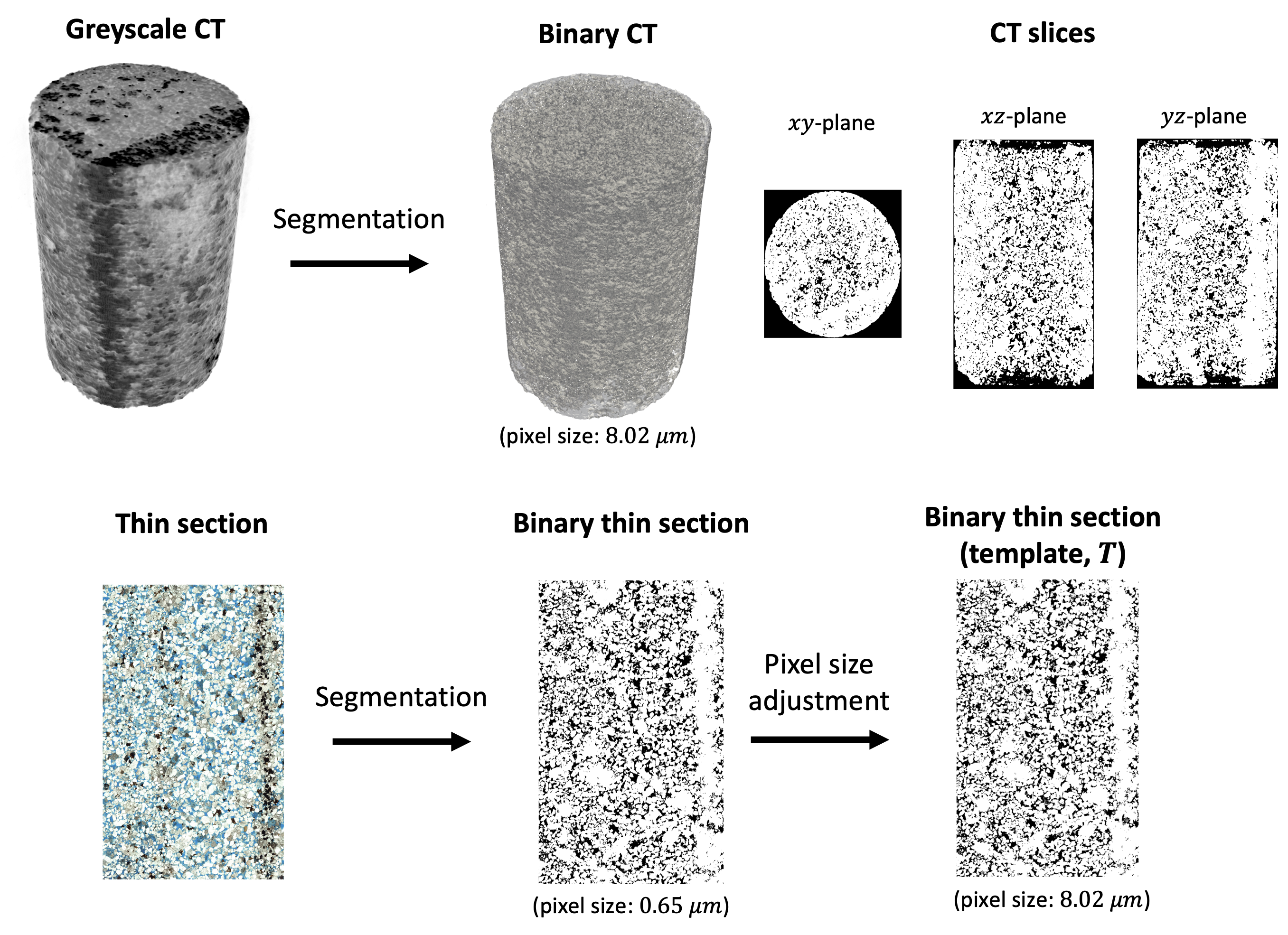}
    \caption{Preprocessing steps, binarization for CT and thin section images to make them in the same feature space (solid (1) and pore (0)). (Top row) The original grayscale CT scan, its segmented binary version, and extracted middle slices along the xy-, yz-, and xz-planes. (Bottom row) The original thin-section image, its segmented binary version, and pixel size adjustment from $0.65 \, \mu m$ to $8.02 \, \mu m$}
    \label{fig:ct_ts_preprocessing}
\end{figure}

We applied the image registration approach to a Berea sandstone digital image. Figure \ref{fig:ct_ts_preprocessing} shows the dataset and preprocessing steps, which convert images from different modalities into the same feature space for image registration. In this study, we segmented binary images (1 for solid and 0 for pore). The top row shows grayscale Berea sandstone CT images ($748 \times 748 \times 1234$) segmented using Otsu’s thresholding \citep{otsu1975threshold} with the commercial software GeoDict \citep{GeoDict2025}. The three rightmost plane views display the segmented CT images from the middle of each axis. The bottom row shows the blue epoxy-impregnated thin section ($13776 \times 8460$), where the epoxy highlights the pores. The thin-section image was segmented using a color-based thresholding approach in the hue-saturation-value (HSV) color space, classifying blue regions as pores and other areas as solid \citep{liu2022rock}. The 3D CT data has a relatively low resolution, with a voxel size of $8.02 \, \mu m$, whereas the voxel size of the thin-section image is $0.65 \, \mu m$. To enable cross-correlation analysis, the thin-section image was downsampled by a factor of $0.65 / 8.02$, aligning its voxel size with that of the CT image and ensuring a consistent spatial resolution between the two datasets. To achieve this, we employed cubic spline interpolation, which preserves fine-scale pore structures and minimizes aliasing artifacts during downsampling.

\begin{figure}[htbp]
    \centering
    \includegraphics[width=0.5\linewidth]{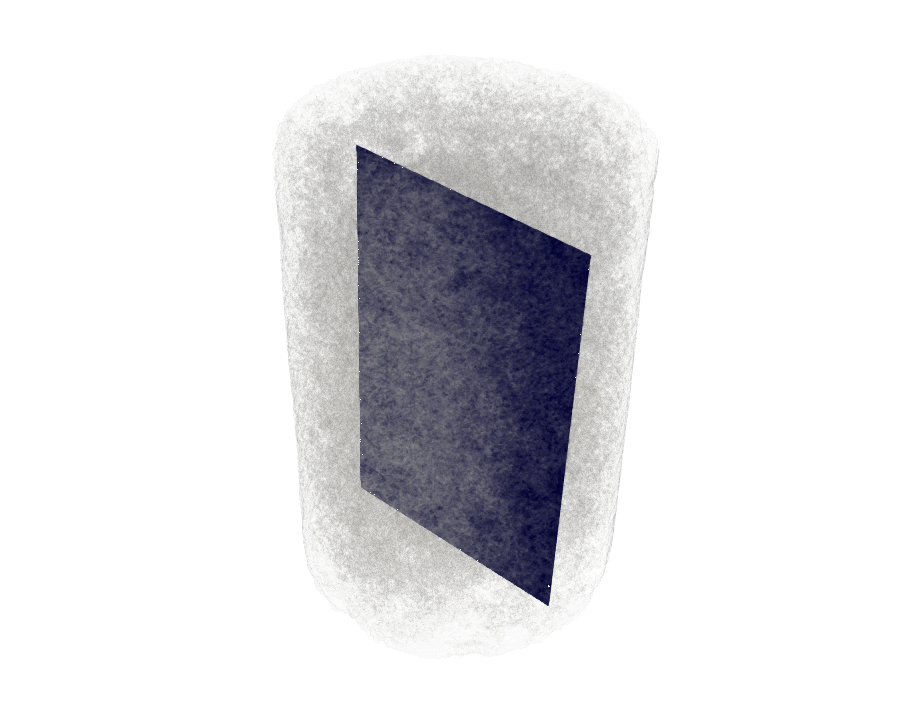}
    \caption{Registered thin-section image within the 3D CT volume of Berea sandstone.}
    \label{fig:B1M1_w_registered_img}
\end{figure}

Since the exact location of the thin section was unknown (i.e., no ground truth), we defined the optimization parameter bounds based on observations of both images. In particular, we specified the x- and y-rotation angle bounds to $(-3, 3)$ degrees, the z-rotation angle bounds to $(0, 10)$ degrees, and the section number range to $1/4 - 3/4$ of the cubic size. Figure \ref{fig:B1M1_w_registered_img} shows the registered thin-section image within the CT volume after optimization. The results indicate that the highest correlation value occurs at section 367, with estimated rotation angles of $0.769^\circ$ for the x-axis, $-1.276^\circ$ for the y-axis, and $7.056^\circ$ for the z-axis.

\begin{figure}[htbp]
    \centering
    \includegraphics[width=1.0\linewidth]{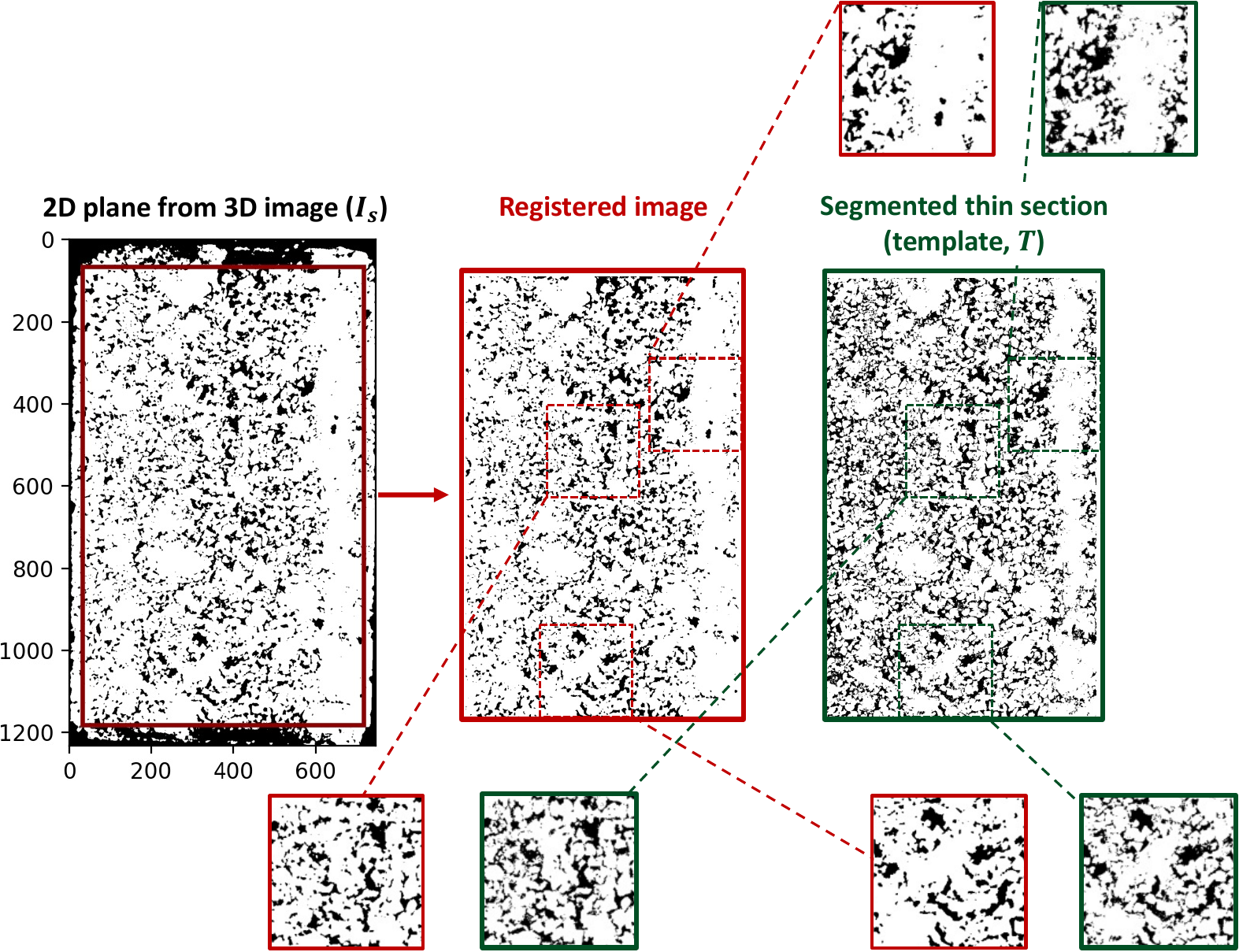}
    \caption{Berea sandstone application: comparison between the registered image extracted from the Berea sandstone 3D volume ($I_s$) and the segmented thin-section template ($T$)}
    \label{fig:real_superimposed}
\end{figure}

\begin{figure}[htbp]
    \centering
    \includegraphics[width=0.5\linewidth]{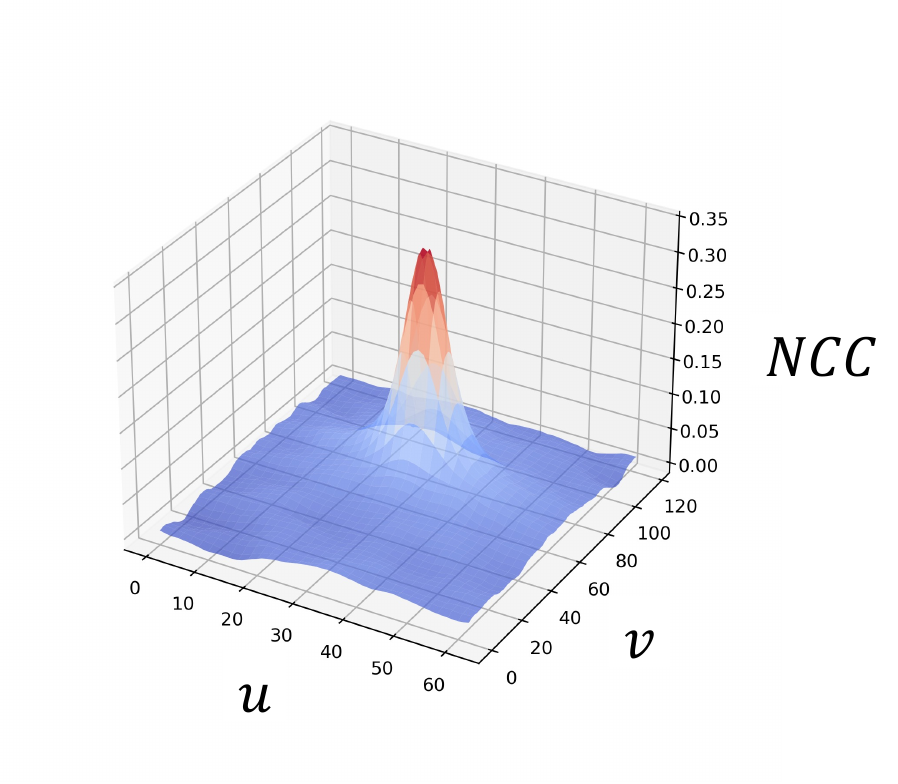}
    \caption{Berea sandstone application: normalized correlation coefficient maps between the extracted 2D plane ($I_s$) and the thin section template ($T$), corresponding to Figure \ref{fig:real_superimposed}}
    \label{fig:3D_correlation_map_real}
\end{figure}

Figure \ref{fig:real_superimposed} shows the extracted 2D plane from the CT volume ($I_s$), the registered image in that plane, and the thin section template. Figure \ref{fig:3D_correlation_map_real} shows the the cross-correlation map between the $I_s$ and the thin section template ($T$) . The correlation map shows a distinct peak at the registered location, with a peak correlation coefficient of $0.35$. To further quantify structural similarity, the structural similarity index measure (SSIM) was computed using the equation \citep{wang2004image, wang2009mean}: 

\begin{equation}
    \text{SSIM}(u, v) = \frac{\left(2\mu_u \mu_v + c_1\right)\left(2\sigma_{uv}+c_2\right)}{\left(\mu_u^2+\mu_y^2+c_1\right)\left(\sigma_u^2+\sigma_v^2+c_2\right)},
\end{equation}
where $\mu_u$ and $\mu_v$ are the mean intensities of the patches, $\sigma_u^2$ and $\sigma_v^2$ are the variance of the patches, $\sigma_{uv}$ is the covariance between patches, $c_1$ and $c_2$ are small constants to avoid instability when the denominator is close to zero. 

The computed SSIM value of 0.990 indicates that the registered image is structurally similar to the thin-section template, with minor discrepancies. The lower correlation coefficient (0.35) and localized mismatches likely stem from differences in resolution and segmentation between the two images, introducing inherent uncertainty in image registration \citep{simonson2006statistics, sykes2009investigation}. However, both qualitative and quantitative comparisons confirm that the microstructures in the registered image and the thin section are well aligned.

\section{Discussions}
\label{sec:Discussion}
Assuming valid registration of the TS image within the CT volume, we explore differences in effective properties and their implications for improving digital rock property estimations by integrating two complementary imaging modalities. The key comparisons are summarized in Table \ref{tab:comparison_2D_images}. While this analysis is limited to 2D image comparisons, preventing direct inference of 3D effective properties, previous studies have shown that 2D-derived properties—including porosity, permeability, and elastic moduli—correlate well with their 3D counterparts when appropriate scaling factors are applied. Readers interested in the relationship between 2D and 3D digital rock properties may refer to \cite{saxena2016estimating, saxena2017estimating, al2018digital}.

\begin{table}[htbp]
    \centering
    \caption{Comparison of effective properties based on 2D images}
    \begin{tabular}{lcc}
        \hline
        Property & CT registered image & TS registered image \\
        \hline
        Porosity & 0.154 & 0.232 \\
        Mean pore size ($\mu$m) & 18.80 & 7.52 \\
        Bulk modulus (GPa) & 15.09 & 11.25 \\
        Shear modulus (GPa) & 16.39 & 11.36 \\
        \hline
    \end{tabular}
    \label{tab:comparison_2D_images}
\end{table}

\subsection{Porosity and pore size distribution}
\begin{figure}[htbp]
    \centering
    \includegraphics[width=0.5\linewidth]{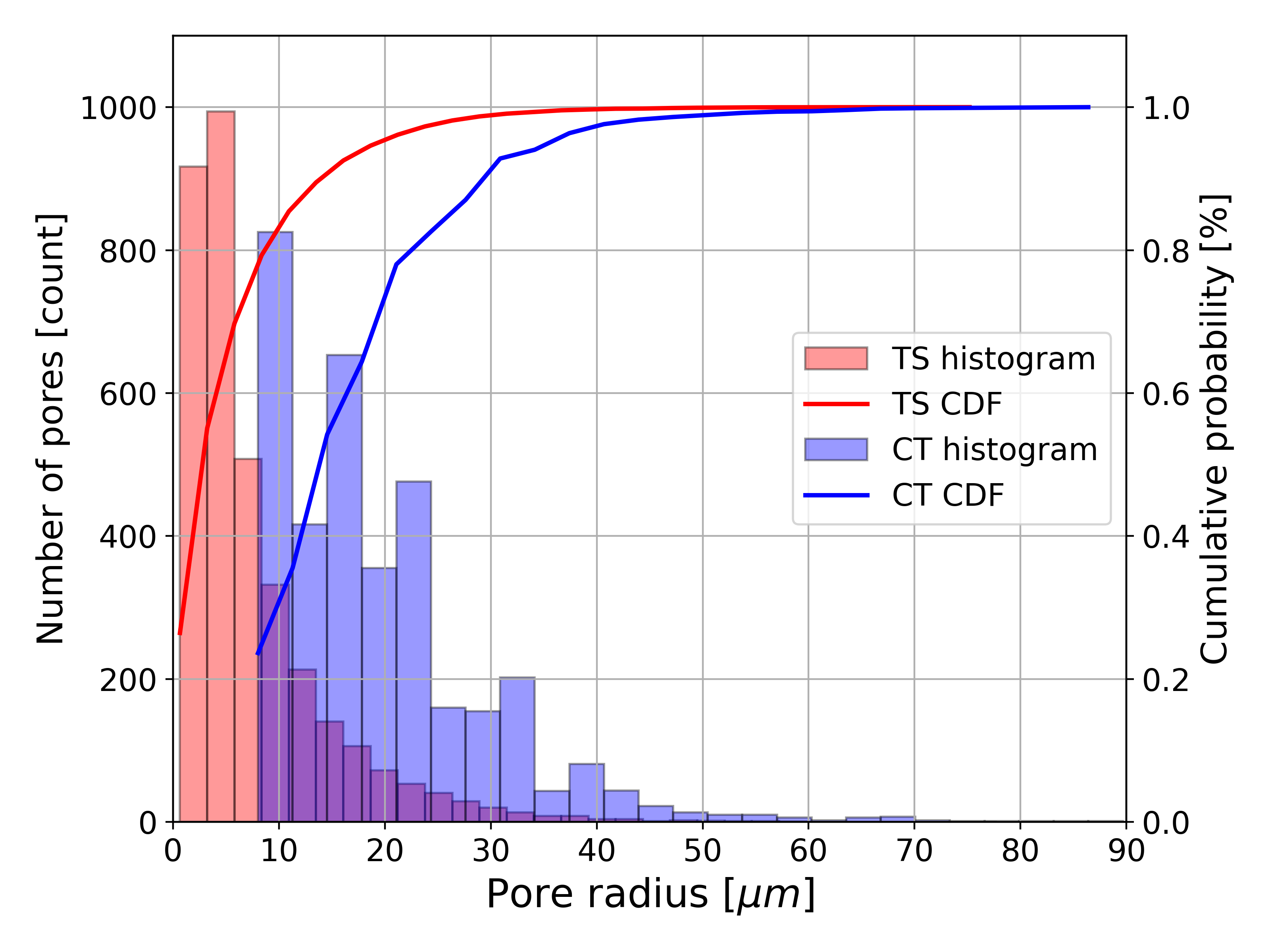}
    \caption{Pore size distribution comparison between the registered CT plane and the thin-section image.}
    \label{fig:pore_size_comparison}
\end{figure}

Pore characteristics play a fundamental role in determining a rock’s effective material properties, including fluid transport \citep{soulaine2016impact, nelson1994permeability}, mechanical behavior \citep{zhang2003pore}, and electrical conductivity \citep{wong1984conductivity}. The registered CT and TS images yield porosities of 0.154 and 0.232, respectively, indicating that the TS image reveals approximately 50 \% more porosity than the CT plane. Figure \ref{fig:pore_size_comparison} shows the pore size distributions obtained from both images. Both exhibit a left-skewed distribution, indicating a higher frequency of smaller pores, as reported in previous studies \citep{andriamihaja2019static, yao2013construction}.

In addition, the TS image reveals thousands of smaller pores with radii below 8.02 $\mu m$ (the CT pixel size). Beyond submicron pore recovery, the TS image also shows a notable reduction in pore sizes compared to the CT image, even for pores larger than the CT resolution limit. These two factors—submicron pore recovery and overall pore size reduction—result in a mean pore radius of 7.52 $\mu m$ in the TS image, compared to 18.80 $\mu m$ in the CT image, meaning pores appear 2.5 times larger in CT than in TS. This suggests that TS images can provide valuable corrections for common biases in CT-derived porosity and permeability estimates, where CT imaging underestimates porosity and overestimates permeability \citep{saxena2019rock, saxena2017effect}.

\subsection{Effective elastic moduli}
Next, we compute the effective elastic moduli of the paired images using an FFT-based solver in GeoDict \citep{GeoDict2025} under a plane strain approximation. Specifically, we assume a two-phase composite, where all solid material is considered quartz, and the remaining regions are pores, based on the primary components of Berea Sandstone \citep{saxena2019rock2}. A nonzero strain of $\varepsilon_{11} = \varepsilon_{22} = 0.001$ and $2\varepsilon_{12} = 0.001$ is applied, along with the physical properties listed in Table \ref{tab:elasticity_properties}. Once the averaged stress ($\sigma$) and strain ($\varepsilon$) are obtained, the bulk and shear moduli of the 2D composite are computed as follows \citep{saxena2016estimating}:

\begin{equation} 
    K_{\text{2D}} = \frac{1}{3} \left(\frac{\sigma_{11} + \sigma_{22} + \sigma_{33}}{\varepsilon_{11} + \varepsilon_{22}}\right) 
\end{equation}

\begin{equation} 
    G_{\text{2D}} = \frac{1}{2} \left(\frac{\sigma_{12}}{\varepsilon_{12}}\right)
\end{equation}

The computed bulk and shear moduli for each 2D image are listed in Table \ref{tab:comparison_2D_images}. The results indicate that the TS-based images yield 25 \% lower bulk moduli and 30 \% lower shear moduli than the CT plane. Although this analysis is limited to 2D comparisons, it highlights the variations in elastic moduli between paired but modality-different digitized images. These findings suggest the potential for correcting 3D image-based effective moduli using numerical simulations, as CT-derived moduli are consistently stiffer than those measured in laboratory experiments \citep{saxena2017effect, saxena2019rock2}.

\begin{table}[htbp]
    \centering
    \caption{Input properties of minerals for elasticity computation \citep{mavko2020rock}}
    \begin{tabular}{lcc}
        \hline
        Phase & Bulk modulus (GPa) & Shear modulus (GPa) \\
        \hline
        Pore & 0 & 0 \\
        Quartz & 36 & 45 \\
        \hline
    \end{tabular}
    \label{tab:elasticity_properties}
\end{table}

\subsection{Geological features}
\begin{figure}[htbp]
    \centering
    \includegraphics[width=0.6\linewidth]{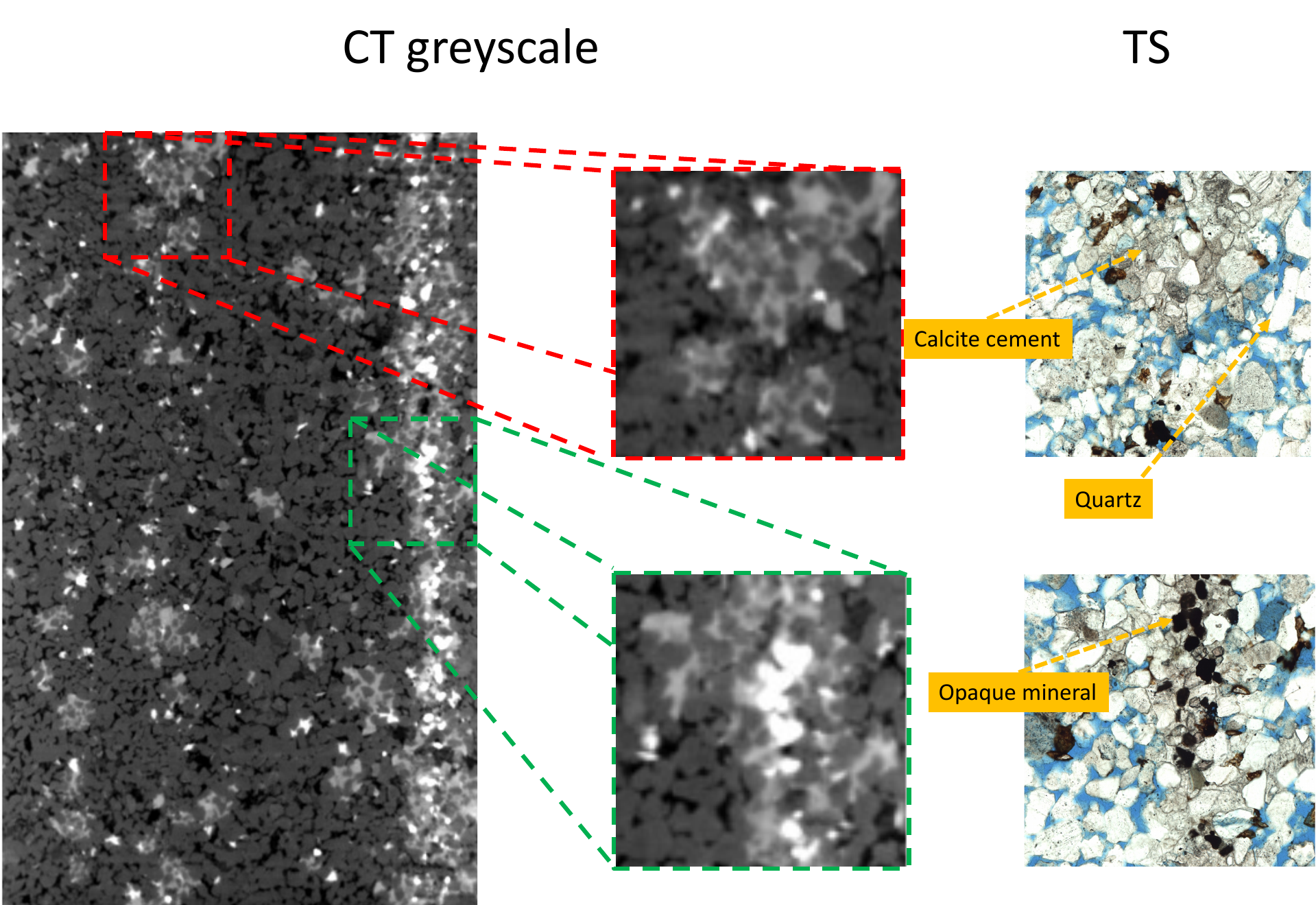}
    \caption{Comparison of paired images: grayscale CT and TS. The TS image provides additional information on cementation and mineralogy that is not clearly visible in the CT plane.}
    \label{fig:CTgrey_TS_comparison}
\end{figure}

Lastly, we examine additional geological insights by directly comparing grayscale CT images with their paired TS images. Figure \ref{fig:CTgrey_TS_comparison} shows the registered grayscale CT image alongside its corresponding TS image for one-to-one comparison. The TS image enhances the visibility of individual grains, which are less distinct in the CT plane, and provides information on calcite cementation. Although Fe-bearing minerals (white) are distinguishable from quartz (dark gray) in our dataset's grayscale CT plane (Figure \ref{fig:CTgrey_TS_comparison}), identifying specific mineralogical information becomes challenging when the 3D digital rock consists of minerals with similar densities (e.g., quartz and orthoclase, which have less than a 3 \% density difference; \citep{goldfarb2022predictive}). Additionally, the TS image reveals weathering effects, distinguishing intact and clean quartz from areas affected by cementation or weathering processes.

By integrating these paired images, we can correlate geological features such as cementation, mineralogy, and weathering conditions. This correlation enables a more accurate assignment of material properties based on real geological characteristics rather than relying solely on intact mineral component properties, thereby improving the accuracy of computed effective properties.

\section{Conclusion}
\label{sec:Conclusion}
We develop a systematic image registration approach to align 2D thin-section images within a 3D CT volume using differential evolution (DE) optimization. The method is first validated on a synthetic porous medium with a known 2D plane location, achieving near-perfect alignment with rotation angle errors below 0.001 degrees. Applying the approach to Berea sandstone, we successfully register the thin-section image within the CT volume, achieving a structural similarity index of 0.990.

With this valid registration, we explore the advantages of paired 2D multimodal images for improving 3D digital rock-based computed effective properties. The thin-section image reveals approximately 50 \% more porosity than the CT plane and shows a consistent reduction in pore sizes, both of which may help correct porosity underestimation and permeability overestimation in 3D CT-based computations. Additionally, bulk and shear moduli computed from thin sections are 25 \% and 30\% lower, respectively, than those derived from CT images, suggesting potential corrections for CT-derived moduli, which tend to be stiffer than laboratory measurements. Beyond these quantitative differences, the thin-section image provides additional geological insights, revealing cementation patterns, mineral phases, and weathering effects—features not discernible in CT images alone. Integrating multimodal imaging improves material property assignment based on actual geological characteristics.

This study demonstrates the potential of multimodal image registration to enhance digital rock-based effective property computations. The proposed method also lays the foundation for data-driven super-resolution techniques and improved numerical simulations leveraging multimodal imaging. Future work could extend this approach to 3D reconstructions using machine learning, further enhancing the predictive capabilities of digital rock physics by incorporating high-resolution thin-section information into 3D models.

\section*{Data Availability Statement}
The code and data used in this study are freely available and can be accessed on the GitHub repository at \url{https://github.com/jh-chung1/ImgRegister2Dto3D}.

\section*{Acknowledgments}
We acknowledge Shell for financial support and for providing the digital rock images. In particular, we appreciate Dr. Ronny Hofmann for sharing insights into the challenges of image registration in practice and industry. This study was primarily conducted using the Sherlock cluster at Stanford University. We are grateful to Stanford University and the Stanford Research Computing Center for providing computational resources and support for this research. We also acknowledge the sponsors of the Stanford Center for Earth Resources Forecasting (SCERF).

\section*{Authorship Statement}
\textbf{Jaehong Chung}: Conceptualization, Investigation, Data Curation, Original Draft, Visualization, Review and Editing.  
\textbf{Tapan Mukerji}: Conceptualization, Investigation, Funding Acquisition, Review and Editing.  
\textbf{Wei Cai}: Conceptualization, Investigation, Funding Acquisition, Review and Editing.  

\bibliographystyle{unsrtnat}
\bibliography{references}  

\end{document}